\documentclass[aps,preprintnumbers,floats, prl, twocolumn, longbibliography, nofootinbib]{revtex4-2}
\usepackage{graphicx}
\usepackage{dcolumn}
\usepackage{bm}
 \usepackage{booktabs}
\usepackage[caption=false]{subfig}
\usepackage{amsmath}
\usepackage{comment}
\usepackage{calligra}
\DeclareMathAlphabet{\mathcalligra}{T1}{calligra}{m}{n}
\DeclareFontShape{T1}{calligra}{m}{n}{<->s*[2.2]callig15}{}

\usepackage{amssymb}
\usepackage{tabularx}
\usepackage{booktabs}
\usepackage{multirow}  
\usepackage{xcolor}
\usepackage{slashed}

\usepackage{url}
\usepackage{fontawesome5}
\usepackage{scalerel}
\usepackage{tikz}
\usetikzlibrary{svg.path} 
\definecolor{orcidlogocol}{HTML}{A6CE39}
\tikzset{
  orcidlogo/.pic={
    \fill[orcidlogocol] svg{M256,128c0,70.7-57.3,128-128,128C57.3,256,0,198.7,0,128C0,57.3,57.3,0,128,0C198.7,0,256,57.3,256,128z};
    \fill[white] svg{M86.3,186.2H70.9V79.1h15.4v48.4V186.2z}
                 svg{M108.9,79.1h41.6c39.6,0,57,28.3,57,53.6c0,27.5-21.5,53.6-56.8,53.6h-41.8V79.1z M124.3,172.4h24.5c34.9,0,42.9-26.5,42.9-39.7c0-21.5-13.7-39.7-43.7-39.7h-23.7V172.4z}
                 svg{M88.7,56.8c0,5.5-4.5,10.1-10.1,10.1c-5.6,0-10.1-4.6-10.1-10.1c0-5.6,4.5-10.1,10.1-10.1C84.2,46.7,88.7,51.3,88.7,56.8z};
  }
}

\newcommand\orcidicon[1]{\href{https://orcid.org/#1}{\mbox{\scalerel*{
\begin{tikzpicture}[yscale=-1,transform shape]
\pic{orcidlogo};
\end{tikzpicture}
}{|}}}}

\definecolor{mycolor}{RGB}{0,0,204}
\definecolor{pink}{RGB}{255,0,127}
\usepackage[unicode]{hyperref}
\hypersetup{
  colorlinks   = true,    
  urlcolor     = blue,    
  linkcolor    =pink,    
  citecolor    = purple      
}

\usepackage{tabularx}
\usepackage{graphicx}
\usepackage{dcolumn}
\usepackage{bm}

\usepackage{float}

\begin{document}

\title{Parametric-Resonance Production of QCD Axions}
 \author{Pirzada \orcidicon{0009-0002-2274-9218}}%
 \affiliation{CAS Key Laboratory of Theoretical Physics, Institute of Theoretical Physics, Chinese Academy of Sciences, Beijing 100190, China}
 \affiliation{School of Physical Sciences, University of Chinese Academy of Sciences, 19A Yuquan Road, Beijing 100049, China}%

\author{Yu Gao  \orcidicon{0000-0002-8228-9981}}%
 \email{gaoyu@ihep.ac.cn }%
 \affiliation{
State Key Laboratory of Particle Astrophysics,
Institute of High Energy Physics, 
Chinese Academy of Sciences,
19B Yuquan Road, Beijing 100049, China}%

\author{Qiaoli Yang \orcidicon{0000-0002-9642-9033}}%
\email{qiaoliyang@jnu.edu.cn}%
\affiliation{Physics Department, College of Physics and Optoelectronic Engineering, Jinan University, Guangzhou 510632, China}%

\begin{abstract}

Dark matter axion production can be significantly enhanced through a generic cosmological mechanism: primordial temperature fluctuations periodically modulate the axion mass during the QCD phase transition, thereby triggering parametric resonance in axion field evolution. This interplay between the resonance and the misalignment mechanism moves the predicted axion mass window for the observed dark matter abundance to $10^{-4}-10^{-3} \, \text{eV}$, shifting the preferred mass to previously unexplored higher ranges. 
\end{abstract}

\maketitle	

\textit{Introduction.} Observations from galaxy dynamics, precision cosmic microwave background (CMB) measurements, and large-scale structures establish that dark matter constitutes $27\%$ of the cosmic energy budget~\cite{Persic:1995ru,Planck:2018vyg}. However, the fundamental nature of dark matter remains elusive. Among proposed candidates, the QCD axion is particularly well-motivated: it was introduced to resolve the strong CP problem by breaking a global $U(1)$ Peccei-Quinn symmetry that dynamically relaxes the QCD $\theta$-angle~\cite{Peccei:1977hh, Peccei:1977ur, Weinberg:1977ma}, with current benchmark models including the KSVZ~\cite{Kim:1979if, Shifman:1979if} and DFSZ~\cite{Dine:1981rt, Zhitnitsky:1980tq} axions. In standard cosmology, QCD axion dark matter is predominantly produced via the misalignment scenario~\cite{Preskill:1982cy, Abbott:1982af, Dine:1982ah}. This framework generally predicts a QCD axion mass around $10^{-5}\,\text{eV}$. If the axion field existed prior to inflation, its quantum fluctuations would imprint isocurvature perturbations in the CMB, leading to stringent observational constraints~\cite{Hertzberg:2008wr, Hamann:2009yf, Wantz:2009it,Beltran:2006sq,Ijaz:2023cvc,Ijaz:2024zma}. However, the prediction depends on the cosmological history and the initial misalignment angle and the current experimental searches have not yet observed a signal in this region. It is therefore important to explore additional cosmological processes that could modify the axion relic abundance and alter the preferred mass range.

A complementary process for generating axion dark matter in the early Universe was proposed in~\cite{Zheng:2025wga}. Primordial density fluctuations from inflation induce periodic temperature variations during the axion mass transition. These fluctuations modulate the axion mass and can trigger parametric resonance, described by instability bands in Mathieu-type dynamics~\cite{1968Theory,Kofman:1997yn,Zheng:2025wga}. Although the primordial temperature perturbations are small, the large number of axion oscillations during the QCD transition allows weak periodic modulation to accumulate (see Fig.~\ref{fig:bands_full_equation}). Unlike previous studies of resonant axion excitation that redistribute an existing axion population~\cite{Sikivie:2021trt}, this process parametrically creates axion modes sourced by primordial temperature fluctuations, even when the initial field displacement is small. Since it is sourced by temperature fluctuations rather than an initial condition, the scenario evades the isocurvature constraints~\cite{Planck:2018jri}, contributing to the relic density alongside the homogeneous misalignment mechanism. 

In this letter, we present a comprehensive study of the mechanism~\cite{Zheng:2025wga} in a realistic cosmological setting and quantify its impact on the QCD axion relic abundance. We compute the field evolution through the resonance epoch, numerically solve the associated mode equation, and quantify the resulting correction to the relic abundance and the corresponding shift in the phenomenologically viable axion mass range. As a result, the predicted dark matter QCD axion mass ($m_a \sim 10^{-5}\,\text{eV}$) can shift toward higher values, motivating future experiments to probe higher-frequency ranges~\cite{Pralavorio:2024neq,yk7n-lrpj,Nguyen:2025ujp,BREAD:2023xhc,Ahmad:2025}.

\medskip
\textit{Field dynamics.}\label{S2} The axion field $\phi(\vec{x},t)$ obeys the Klein--Gordon equation:
\begin{equation}\label{eq1}
D^{\mu}\partial_{\mu}\phi(\vec{x},t)-m^{2}(T)\,\phi(\vec{x},t)=0~,
\end{equation}
where $D^\mu$ is the covariant derivative and $m(T)$ is the temperature-dependent axion mass that evolves with the cosmological temperature $T(t)$. To incorporate leading-order scalar perturbations, we work in the conformal Newtonian gauge and retain terms linear in the metric potentials $\Phi$ and $\Psi$. The resulting E.O.M becomes 
\begin{equation}
\label{eq2341_std}
\ddot\phi+3H\dot\phi-\frac{1}{a^{2}}\nabla^{2}\phi+m^{2}(T)\,\phi
+f(\vec{x},t,\phi)=0~.
\end{equation}
The function $f(\vec{x},t,\phi)$ encodes scalar perturbations. It includes: (i) renormalization of the kinetic operator, (ii) a modified friction term, (iii) gradient effects from the perturbed spatial metric, and (iv) modulation of the effective mass (see supplementary material for derivations). Explicitly: 
\begin{equation}
\begin{aligned}
f(\vec x,t,\phi)=&-2\Psi\,\ddot\phi-\left(\dot\Psi-3\dot\Phi+6H\Psi\right)\dot\phi
+\frac{2\Phi}{a^2}\nabla^{2}\phi\\
&-\frac{1}{a^2}\partial_j(\Phi+\Psi)\,\partial_j\phi
+\frac{{\rm d}m^{2}}{{\rm d}T}\,\delta T\,\phi~.
\end{aligned}
\end{equation}
The resonance parameters are time-dependent, causing the instability band to sweep through momentum space. The system is driven by a stochastic primordial potential. To capture the full dynamics, we solve the mode equation numerically and perform an ensemble average over realizations of the primordial perturbation.

Eq. \ref{eq2341_std} represents the canonical setup for parametric resonance. It describes a damped oscillator with a time-dependent frequency. The periodic mass modulation originates from scalar metric perturbations. During the radiation dominated era, the scale factor evolves as $a(t) = (t/t_1)^{1/2}$, and the comoving wavenumber scales as $k^2/a^2(t) = k^2 t_1/t$, where $t_1$ denotes the time when oscillations begin. The unperturbed, homogeneous $\Phi$ equation is 
\begin{equation}
\ddot\phi(\vec{k},t)+\frac{3}{2t}\dot\phi(\vec{k},t)+\omega_k^2(t)\,\phi(\vec{k},t)=0
\label{eq_omegadef}
\end{equation}
with $\omega_k^2(t) \equiv k^2 t_1/t + m^2(t)$. Resonance effects arise from the terms in $f(\vec{x},t,\phi)$. Substituting the sub-horizon radiation-era solution for $\Phi(\vec{k},t)$ and the temperature perturbation $\delta T/T$~\cite{Dodelson:2003ft}. Because the resonance parameter is proportional to the small primordial perturbation amplitude and each mode traverses the narrow instability band only briefly as the axion mass evolves, the net amplification during each crossing remains modest and the dynamics stays in the linear regime, allowing nonlinear backreaction and mode coupling to be neglected. Thus it yields the $k$-space equation; see the Supplementary Material and Refs.~\cite{kaya,MukhanovEtAl1992,MaBertschinger1995,Fauth,Bracewell1966TheFT}:
\begin{equation}\label{foeq}
\begin{aligned}
&\Bigg[1-\frac{9\,\Phi_p(\vec{k})}{2k^2\,t\,t_1}\cos\Theta_k(t)\Bigg]\ddot{\phi}(\vec{k},t)+
\\[2pt]
&\Bigg[\frac{3}{2t}
+\frac{9\,\Phi_p(\vec{k})}{\sqrt{3}\,k\,t^2}\sqrt{\frac{t}{t_1}}\sin\Theta_k(t)
-\frac{27\,\Phi_p(\vec{k})}{4k^2\,t^2\,t_1}\cos\Theta_k(t)\Bigg]\dot{\phi}(\vec{k},t)
\\[2pt]
&+\Bigg[k^2\frac{t_1}{t}+m^2(t)
+\Bigg(\frac{9\,\Phi_p(\vec{k})}{2t^2}-\frac{3\,\Phi_p(\vec{k})}{2}\frac{{\rm d}m^2(t)}{{\rm d}T}\,T(t)\Bigg)\\&\times\cos\Theta_k(t)\Bigg]\phi(\vec{k},t)
=0~,
\end{aligned}
\end{equation} 
Here, $a(t_1) = 1$ is adopted for simplicity. The phase $\Theta_k(t)\equiv2k\sqrt{t_1 t/3}$ arises from acoustic oscillations of the radiation fluid potential $\Phi\propto \cos(k\eta/\sqrt{3})$, where conformal time $\eta$ is converted to cosmic time via $\eta(t) = \int^t dt'/a(t') = 2\sqrt{t t_1}$. The time $t_1$ marks the onset of axion oscillations and is implicitly defined by  
\begin{equation}\label{axionmasstimess}
\begin{aligned}
t_1\simeq\left({t_2^{2n}}/{m^2}\right)^{\frac{1}{2n+2}}~,
\end{aligned}
\end{equation}
with the QCD transition time $t_2 \sim 10^{-5}\,\text{s}$ at a temperature $T\sim 100\,\text{MeV}$~\cite{Husdal:2016haj}.
For the mass-temperature we adopt the powerlaw relation $m_a(T)\propto T^{-n}$ and take $n=3$ as given by lattice results~\cite{Borsanyi:2016nature20115, Petreczky:2016TopSuscepAxionCosmo, Athenodorou:2022JHEP10_197, Chen:2022PhysRevD106_074501, Kotov:2025JHEP09_045}. Here the dominant uncertainty in the resonance efficiency arises, since the Mathieu parameter $q$ scales linearly with $n$. In the linear-response regime, this implies $\delta\rho_1/\rho_1 \simeq 2\delta n/n$, up to additional small corrections associated with shifts in the resonance time. The preferred axion masses shown in Fig.~3 therefore should be regarded as benchmark values corresponding to the lattice-motivated choice $n=3$. For example, changing the exponent to $n=2.5$ results in an approximately 34\% change in the resonance-generated density.

Eq. \ref{foeq} reveals three distinct modulation channels: (i) the $\ddot{\phi}$ term renormalizes effective inertia, (ii) the $\dot{\phi}$ term modulates friction, and (iii) the remaining terms involve mass modulation via $({\rm d}m^2/{\rm d}T)T$, sourced by $\delta T$. As $m(t)$ increases during the QCD phase transition, this mass modulation dominates energy injection. While $m(t)$ introduces nontrivial time dependence, useful approximations emerge by isolating dominant terms and recasting the equation into Mathieu-type form. 

\medskip
{\it Parametric resonance} 
is most efficient when the periodic drive is near twice the instantaneous frequency
(e.g., the $l=2$ band), or more generally near integer harmonics:  
\begin{equation}
\omega_{\rm drive}(t)\simeq l\,\omega_k(t),\qquad l=1,2,3,\dots~
\label{eq_rescond_general}
\end{equation}
The driving frequency is determined by the phase in Eq.~\ref{foeq}. The dominant ($l=2$) resonance condition $\omega_{\rm drive} \simeq 2\omega_k$ implies a short-lived resonance event for each $k$-mode:  
\begin{equation}
\begin{aligned}
\label{111321}
t_R=\begin{cases}
 t_2\left(\frac{k}{k_{2} } \right)^{\frac{2}{2n+1} }~~~~\text{for}~k< k_2~,\\
t_2\left(\frac{k}{k_{2} } \right)^{2}~~~~~~~~\text{for}~k> k_2~,
\end{cases}
\end{aligned}
\end{equation}
depending on whether the resonance occurs during the growth phase $t_1 \le t \le t_2$ or after mass growth saturates. Here, $k_2$ denotes a reference scale: $k_2 \equiv m \sqrt{{3t_2}/{t_1}}$. 

\begin{figure}[ht!]
\centering
\includegraphics[width=0.47\textwidth]{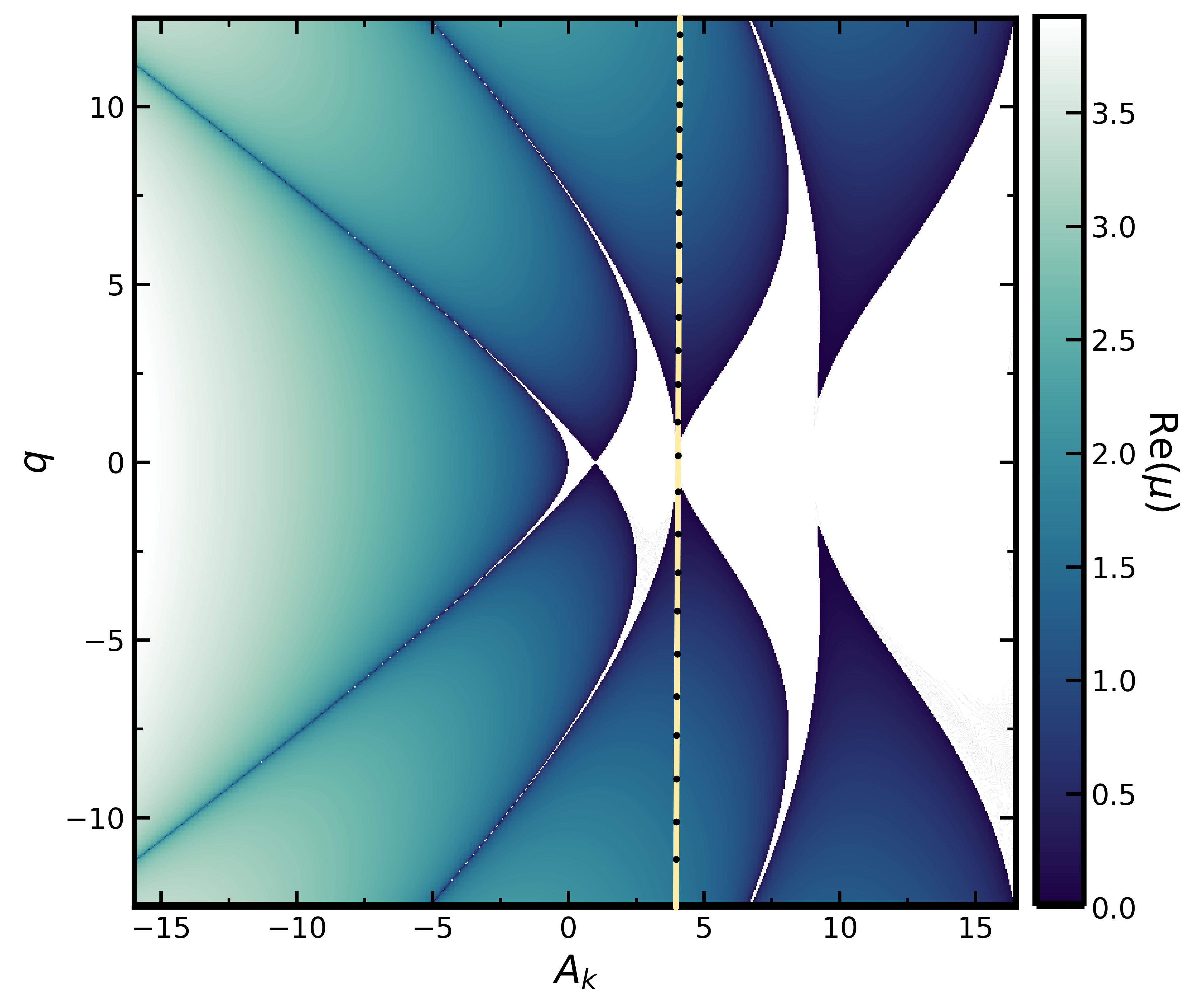}
\caption{Mathieu instability chart showing the largest Floquet exponent $\mathrm{Re}(\mu)$ in the $(A_k,q)$ plane. Colored regions indicate parametric resonance ($\mathrm{Re}(\mu)>0$); the black contour outlines the instability boundary, and the vertical line marks the $l=2$ resonance center at $A_k=4$.}
\label{fig:mathieu_instability}
\end{figure}

During the brief interval around $t \simeq t_R(k)$, the coefficients in Eq.~\ref{foeq} vary slowly compared to the oscillatory phase. Neglecting Hubble friction and cosmic expansion over this narrow window, the mode equation reduces to a Mathieu-type form by introducing the rescaled time variable: 
\begin{equation}
z\equiv \frac{k}{\sqrt{3}}\,\eta(t)=2k\sqrt{\frac{t t_1}{3}}~,
\label{eq_zdef}
\end{equation}
which transforms oscillatory terms into sinusoidal dependence on $z$. After a field redefinition to remove the leading first-derivative term, the mode equation becomes canonical (see supplementary material):  
\begin{equation}
\partial^2_z\phi_{k} + \left( A_{k} - 2q \cos 2z \right) \phi_{k} = 0~,
\label{eq_mathieu}
\end{equation}
with Mathieu parameters: 
\begin{equation}
A_k=3+\frac{3m^2(t)}{k^2}~,\ \ 
q= \frac{{\rm d}m^2(t)}{{\rm d}T}\frac{9\Phi_{p}(\vec{k})T(t)}{4k^2}~. \label{eq:matheiu_parameters}
\end{equation}

The Mathieu parameters $A_k$ and $q$ provide a local approximation near individual resonance episodes. Since the coefficients in Eq.~\ref{foeq} vary with time, a given mode can encounter multiple short resonance episodes. The net enhancement therefore cannot be associated with a single Floquet exponent; rather, it is determined by solving the full time-dependent equation and evaluating $\rho_1$ after the resonance epoch. Instability bands occur when $A_k \simeq l^2$ for integer $l$. The dominant $l=2$ band corresponds to $A_k \simeq 4$, and for small parameter $q$, the band width scales as $\Delta A_k \sim q^2$. The growth rate $\mu(t) \equiv {\rm d}\ln|\phi_k|/{\rm d}t$ illustrates the resonance behavior, as shown in Fig.~\ref{fig:mathieu_instability}.

\begin{figure}[ht!]
\centering
\includegraphics[width=0.47\textwidth]{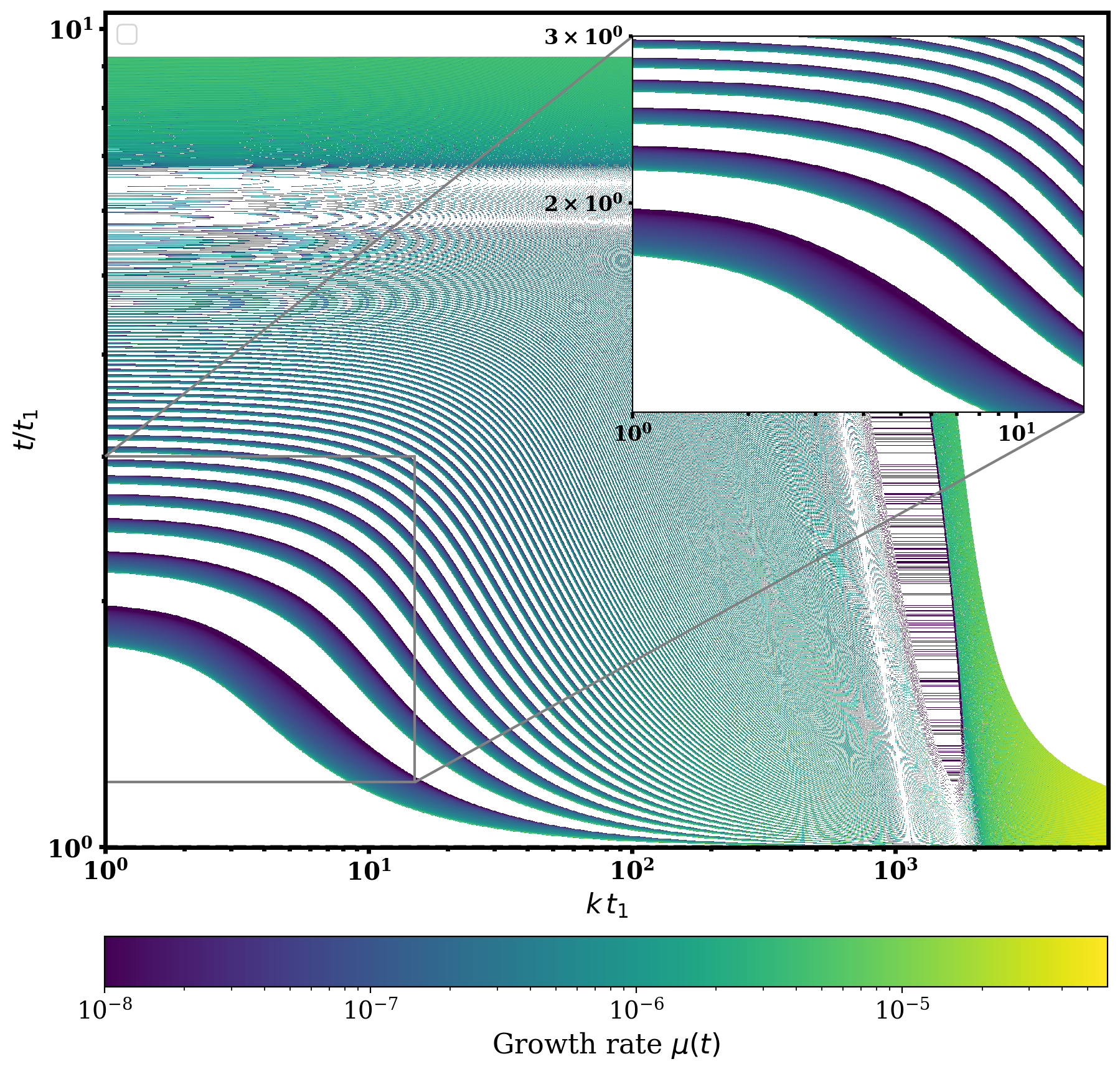}
\caption{Resonance band-map computed from the full equation of motion. The instantaneous growth rate $\mu(\tau,\kappa) \equiv t_1^{-1}\,{\rm d}\ln|\phi(\tau;\kappa)|/{\rm d}\tau$ is derived from numerical solutions of Eq.~\ref{foeq}. Stable regions between bands are left blank. Band structures widen and become more visible in the weak-driving regime (lower-left). Broken patterns near $2 \times 10^{3}$ reflect numerical precision limits.}

\label{fig:bands_full_equation}
\end{figure}

Full dynamics exhibits richer structure than the qualitative discussion with the narrow-resonance Mathieu analysis above. The simulations reveal recurring short-time resonant behavior when Eq.~\ref{foeq} is solved numerically. The field growth rate is plotted over the dimensionless $(\kappa,\tau)$ plane Fig.~\ref{fig:bands_full_equation}, where $\kappa \equiv k t_1$ and $\tau \equiv t/t_1$. The resulting instability bands pinpoint epochs where periodic metric-sourcing terms efficiently produce axions. A single $k$-mode typically encounters multiple resonance bands during the phase transition. The locations of these bands drift over time. The drift arises because both the effective frequency $\omega_k^2(\tau) = k^2/a^2 + m^2(\tau)$ and the driving amplitude evolve across the transition near $t \simeq t_2$. For larger $k$, parametric drive dominates growth. Resonant bands progressively narrow and densely cluster as $k$ increases, with growing structural complexity.

\medskip
{\it Relic density} of the resonance-enhanced growth is calculated through numerical integration of Eq.~\ref{foeq}, which we then compare to the homogeneous case. First, a baseline solution $\phi_{\rm b}(t,k)$ for misalignment is computed by solving the homogeneous Eq.~\ref{eq_omegadef} with initial conditions $\phi_{\rm b}(t_1,k)=\theta_0 f_a,$ and $ \dot{\phi}_{\rm b}(t_1,k)=0\,$,
where $\theta_0$ denotes the initial misalignment angle. Second, we isolate the resonance-generated component as $\phi_1(t,k;\Phi_p)\equiv \phi_{\rm_f}(t,k;\Phi_p)-\phi_{\rm_b}(t,k)\,$, ensuring both $\phi_1$ and $\dot{\phi}_1$ vanish at $t_1$. The system is driven by primordial fluctuations with a nearly scale-invariant power spectrum~\cite{Planck:2018jri}:  
\begin{equation}
\mathcal{P}_\Phi(k)\equiv \frac{k^3}{2\pi^2}\,\langle |\Phi_{\rm }(\bm{k})|^2\rangle\,.
\label{eq:Pphi_def_num}
\end{equation}
We restrict to the linear regime and adopt a $k$-independent amplitude~\cite{Sikivie:2021trt}   $\langle |\Phi_p(k)|^2\rangle=\mathcal{P}_\Phi(k)\simeq A\,$, with $A\simeq \mathcal{O}(10^{-9})$ \cite{Naess:2025DR6maps}. 

Since Eq.~\ref{foeq} is linear in $\Phi_p$, the resonance-generated field satisfies $\phi_1(t,k;\Phi_p) \propto \Phi_p$, leading to axion energy densities scaling quadratically with $\Phi_p$. Our numerical procedure solves the equations for fixed realizations of $\Phi_p$ and averages over Gaussian-distributed $\Phi_p$ with $\langle \Phi_p \rangle = 0$ and $\mathrm{Var}(\Phi_p)=A$. The energy density for each mode is
\begin{equation}
E_1(t,k;\Phi_p)=\frac{1}{2}\dot{\phi}_1^2(t,k;\Phi_p)
+\frac{1}{2}\,\omega_k^2(t)\,\phi_1^2(t,k;\Phi_p)\,.
\label{eq:E1_num}
\end{equation}
The ensemble-averaged spectral energy density per logarithmic interval is
\begin{equation}
\frac{{\rm d}\rho_1}{{\rm d}\ln k}(t,k)\equiv \left\langle E_1(t,k;\Phi_p)\right\rangle_{\Phi_p}\,.
\label{eq:drho_dlnk_num}
\end{equation}
The resonance-generated component remains in the linear-response regime, with $\phi_1 \propto \Phi_p$ and $\rho_1\propto\langle\Phi_p^2\rangle=A$. This scaling demonstrates that the relic-density correction originates from the physical resonance response rather than from numerical transients. Numerical stability has been verified by varying the time-step size, integration interval, and resonance-band resolution. The resulting relic abundance changes by less than 1\% in all cases, indicating that the results are insensitive to phase-coherence artifacts, transient-band resolution, and numerical noise at the precision level of the present analysis. The total energy density is obtained by integrating over resonant modes $k \in [t^{-1}_1, k_2]$, as defined in Eq.~\ref{111321}, to set a conservative upper bound. As shown in Fig.~\ref{fig:bands_full_equation}, higher-$k$ modes resonate earlier and experience greater redshift by $t_2$. In simulations, the majority of resonance-enhanced modes become nonrelativistic by the end of the QCD phase transition.

\begin{figure*}[t]
\centering
\includegraphics[width=0.99\textwidth,height=0.25\textheight]{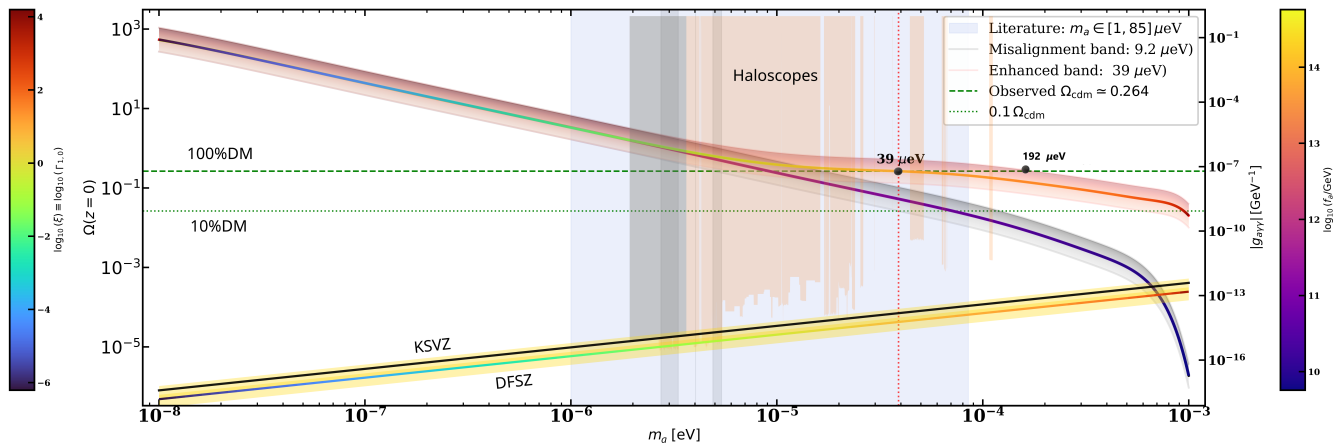}
\caption{Impact of parametric resonance on the QCD axion relic abundance. Homogeneous misalignment (gray) and resonance-enhanced (red) predictions are shown as bands, with widths reflecting the initial misalignment angle $\theta_0 \in [1,2]$. A benchmark case with $\theta_0=1.3$ is highlighted via solid curves through the band centers. Resonance-enhanced scenarios account for 100\% dark matter at axion masses $39~(192)~\mu$eV for $\theta_0=1.3~(2.0)$. For comparison, literature predictions and existing haloscope constraints~\cite{AxionLimits} are overlaid. The lower band (yellow) illustrates the mass-coupling relation across QCD axion models.}

\label{fig:relic_density}
\end{figure*}

Eq. \ref{eq:drho_dlnk_num} still retains spatial gradients that redshift differently from nonrelativistic matter. This component is negligible if boosted modes are nonrelativistic. To quantify today’s relic density, we define a ``diluted ratio" $\Gamma \equiv \tilde{\rho}_1/\rho_0$, where tilde excludes spatial gradient terms in $\rho_1$, and $\rho_1/\rho_0 \to \Gamma$ at low redshift. The present-day dark matter fraction is then  
\begin{equation}
\Omega=(1+\Gamma)~\frac{\rho_0(t_{\rm _*};m)}{\rho_{c,0}}
\left(\frac{a_{\rm _*}}{a_0}\right)^3.
\label{eq:Omega_per_fa2}    
\end{equation}
Here $\rho_{c,0}$ as the present day critical density and $\rho_0$ the misalignment energy density~\cite{Preskill:1982cy, Sikivie:2009qn, Abbott:1982af}. Here, $t_*$ and $a_*$ denote the time and scale factor at the end of mass modulation.

\medskip
\textit{Results \& Discussion.}
\label{S:QCDlineScan}
We compute the parametric resonant equations for an axion mass range $m \in [10^{-8}, 10^{-3}]$ eV. Imposing an upper limit $m \leq 10^{-3}$ eV ensures the parameter $q$ remains small, avoiding excessively large perturbations. Predictions for present-day relic density are illustrated in Fig.~\ref{fig:relic_density}. We assume an ${\cal O}(1)$ initial misalignment angle $\theta_0 \in [1,2]$ to be consistent with unexplored regions in the classical window. While the choice of $\theta_0$ shifts overall predictions for both scenarios, the ratio between resonance-enhanced and homogeneous misalignment abundances remains $\theta_0$-independent. The numerical enhancement can be interpreted as secular accumulation during the time-dependent QCD transition. In the scenario adopted here, the resonance-generated component may be roughly estimated as $\Omega_1\propto A(m_0t_2)^{7/4}$ for $n=3$. Thus, the small primordial amplitude is compensated by the long evolution time $t_2$ relative to the field oscillation timescale $m_0^{-1}$.

The resonant amplification could shift the preferred QCD axion dark matter target above $10^{-5} \,\text{eV}$. If misalignment alone explains the observed dark matter density, the resonance raises the preferred axion mass from $\sim10\,\mu$eV to $39\,\mu$eV and $192\,\mu$eV for $\theta_0 = 1.3$ and $2$, respectively—masses exceeding the sensitivity of current haloscope experiments. These preferred values for 100\% dark matter remain subject to cosmological uncertainties in plasma modeling and primordial perturbations.

For low masses ($m \ll 10\,\mu$eV), a smaller $\theta_0$ is typically assumed. From Eq.~\ref{eq:matheiu_parameters}, the parameter $q \propto {\rm d}m^2/{\rm d}T$, so a smaller axion mass quadratically reduces the pumping efficiency, suppressing resonance effectiveness in this regime. If misalignment accounts for only a fraction of the relic abundance (as between the 10\% and 100\% DM lines in Fig.~\ref{fig:relic_density} ), the resonance provides sufficient density boosts at $10^{-3}-10^{-5}\,\text{eV}$ to meet observational requirements.  

\medskip
To summarize, this Letter shows that mass parametric resonance seeded by primordial perturbations provides an efficient process for cosmic QCD axion generation. This process shifts the preferred QCD axion dark matter mass window upward, because additional axion population can be generated even when the misalignment contribution is small. We present a comprehensive treatment of mass-oscillation-driven field evolution, showing that the resonant band maps predict recurring growth episodes where energy is transferred from the plasma environment into axion populations. Analysis via Mathieu-type simplification during the resonant interval reveals periodic amplification of axion modes at larger axion masses and lower $k$. Since temperature perturbations are an intrinsic feature of the cosmological plasma, this advocates parametric resonance as an inflation-seeded phenomenon for QCD axions and may also apply to other light bosonic dark matter production.

\medskip
\textit{Acknowledgments.}
The authors thank Nadir Ijaz for assistance with numerical implementation. This work was supported in part by the National Natural Science Foundation of China (Grant No. 12447105).
\bibliography{name}

@article{yk7n-lrpj,
  title = {New constraints on dark photon dark matter with a millimeter-wave dielectric haloscope},
  author = {Wei and others},
  journal = {Phys. Rev. Lett.},
  pages = {--},
  year = {2026},
  month = {Jan},
  publisher = {American Physical Society},
  doi = {10.1103/yk7n-lrpj},
  url = {https://link.aps.org/doi/10.1103/yk7n-lrpj}
}

@article{Pralavorio:2024neq,
    author = "Pralavorio, Pascal",
    collaboration = "MADMAX",
    title = "{The axion dark matter experiment MADMAX}",
    eprint = "2409.20169",
    archivePrefix = "arXiv",
    primaryClass = "hep-ex",
    doi = "10.22323/1.476.0751",
    journal = "PoS",
    volume = "ICHEP2024",
    pages = "751",
    year = "2025"
}

@article{Nguyen:2025ujp,
    author = "Nguyen, Le Hoang and Horns, Dieter and Lobanov, Andrei",
    title = "{Permanently magnetized axion-photon conversion surface for direct dark matter searches with BRASS-p}",
    eprint = "2509.20583",
    archivePrefix = "arXiv",
    primaryClass = "physics.ins-det",
    doi = "10.1088/1748-0221/20/12/P12026",
    journal = "JINST",
    volume = "20",
    number = "12",
    pages = "P12026",
    year = "2025"
}

@article{BREAD:2023xhc,
    author = "Knirck, Stefan and others",
    collaboration = "BREAD",
    title = "{First Results from a Broadband Search for Dark Photon Dark Matter in the 44 to 52{\,}{\,}{\ensuremath{\mu}}eV Range with a Coaxial Dish Antenna}",
    eprint = "2310.13891",
    archivePrefix = "arXiv",
    primaryClass = "hep-ex",
    reportNumber = "FERMILAB-PUB-23-625-PPD",
    doi = "10.1103/PhysRevLett.132.131004",
    journal = "Phys. Rev. Lett.",
    volume = "132",
    number = "13",
    pages = "131004",
    year = "2024"
}

@article{Sikivie:2009qn,
  author  = {Sikivie, P. and Yang, Q.},
  title   = {Bose-Einstein condensation of dark matter axions},
  journal = {Phys. Rev. Lett.},
  volume  = {103},
  pages   = {111301},
  year    = {2009},
  doi     = {10.1103/PhysRevLett.103.111301}
}

@article{Peccei:1977hh,
  author  = {Peccei, R. D. and Quinn, H. R.},
  title   = {CP Conservation in the Presence of Pseudoparticles},
  journal = {Phys. Rev. Lett.},
  volume  = {38},
  pages   = {1440},
  year    = {1977},
  doi     = {10.1103/PhysRevLett.38.1440}
}

@article{Peccei:1977ur,
  author  = {Peccei, R. D. and Quinn, H. R.},
  title   = {Constraints Imposed by CP Conservation in the Presence of Instantons},
  journal = {Phys. Rev. D},
  volume  = {16},
  pages   = {1791},
  year    = {1977},
  doi     = {10.1103/PhysRevD.16.1791}
}

@article{Weinberg:1977ma,
  author  = {Weinberg, S.},
  title   = {A New Light Boson?},
  journal = {Phys. Rev. Lett.},
  volume  = {40},
  pages   = {223},
  year    = {1978},
  doi     = {10.1103/PhysRevLett.40.223}
}

@article{Kim:1979if,
  author  = {Kim, J. E.},
  title   = {Weak-Interaction Singlet and Strong CP Invariance},
  journal = {Phys. Rev. Lett.},
  volume  = {43},
  pages   = {103},
  year    = {1979},
  doi     = {10.1103/PhysRevLett.43.103}
}

@article{Shifman:1979if,
  author  = {Shifman, M. A. and Vainshtein, A. I. and Zakharov, V. I.},
  title   = {Can Confinement Ensure Natural CP Invariance of Strong Interactions?},
  journal = {Nucl. Phys. B},
  volume  = {166},
  pages   = {493},
  year    = {1980},
  doi     = {10.1016/0550-3213(80)90209-6}
}

@article{Dine:1981rt,
  author  = {Dine, M. and Fischler, W. and Srednicki, M.},
  title   = {A Simple Solution to the Strong CP Problem with a Harmless Axion},
  journal = {Phys. Lett. B},
  volume  = {104},
  pages   = {199},
  year    = {1981},
  doi     = {10.1016/0370-2693(81)90590-6}
}

@article{Zhitnitsky:1980tq,
  author  = {Zhitnitsky, A. R.},
  title   = {On Possible Suppression of the Axion Hadron Interactions},
  journal = {Sov. J. Nucl. Phys.},
  volume  = {31},
  pages   = {260},
  year    = {1980}
}

@article{Preskill:1982cy,
  author  = {Preskill, J. and Wise, M. B. and Wilczek, F.},
  title   = {Cosmology of the Invisible Axion},
  journal = {Phys. Lett. B},
  volume  = {120},
  pages   = {127},
  year    = {1983},
  doi     = {10.1016/0370-2693(83)90637-8}
}

@article{Abbott:1982af,
  author  = {Abbott, L. F. and Sikivie, P.},
  title   = {A Cosmological Bound on the Invisible Axion},
  journal = {Phys. Lett. B},
  volume  = {120},
  pages   = {133},
  year    = {1983},
  doi     = {10.1016/0370-2693(83)90638-X}
}

@article{Dine:1982ah,
  author  = {Dine, M. and Fischler, W.},
  title   = {The Not-So-Harmless Axion},
  journal = {Phys. Lett. B},
  volume  = {120},
  pages   = {137},
  year    = {1983},
  doi     = {10.1016/0370-2693(83)90639-1}
}

@article{Ijaz:2023cvc,
    author = "Ijaz, Nadir and Mehmood, Maria and Rehman, Mansoor Ur",
    title = "{The stochastic gravitational-wave background from primordial black holes and observable proton decay in R-symmetric SU(5) Inflation}",
    eprint = "2308.14908",
    archivePrefix = "arXiv",
    primaryClass  = "astro-ph.CO",
    doi = "10.1140/epjc/s10052-025-15078-w",
    journal = "Eur. Phys. J. C",
    volume = "85",
    number = "12",
    pages = "1394",
    year = "2025"
}

@inproceedings{Bracewell1966TheFT,
  title={The Fourier Transform and Its Applications},
  author={Ronald N. Bracewell},
  year={1966},
  url={https://api.semanticscholar.org/CorpusID:18010056}
}

@article{Naess:2025DR6maps,
  author       = {Naess, Sigurd and others},
  title        = {The Atacama Cosmology Telescope: DR6 maps},
  journal      = {JCAP},
  year         = {2025},
  month        = {11},
  pages        = {061},
  doi          = {10.1088/1475-7516/2025/11/061},
  eprint       = {2503.14451},
  archivePrefix= {arXiv},
  primaryClass = {astro-ph.CO}
}

@article{Petreczky:2016TopSuscepAxionCosmo,
  author        = {Petreczky, Peter and Sch{\"a}dler, Hans-Peter and Sharma, Sayantan},
  title         = {The topological susceptibility in finite temperature QCD and axion cosmology},
  journal       = {Physics Letters B},
  volume        = {762},
  pages         = {498--505},
  year          = {2016},
  month         = nov,
  doi           = {10.1016/j.physletb.2016.09.063},
  eprint        = {1606.03145},
  archivePrefix = {arXiv},
  primaryClass  = {hep-lat}
}

@article{Ijaz:2024zma,
    author = "Ijaz, Nadir and Rehman, Mansoor Ur",
    title = "{Exploring primordial black holes and gravitational waves with R-symmetric GUT Higgs inflation}",
    eprint = "2402.13924",
    archivePrefix = "arXiv",
    primaryClass  = "astro-ph.CO",
    doi = "10.1016/j.physletb.2024.139229",
    journal = "Phys. Lett. B",
    volume = "861",
    pages = "139229",
    year = "2025"
}

@article{Wantz:2009it,
  author  = {Wantz, O. and Shellard, E. P. S.},
  title   = {Axion cosmology revisited},
  journal = {Phys. Rev. D},
  volume  = {82},
  pages   = {123508},
  year    = {2010},
  doi     = {10.1103/PhysRevD.82.123508}
}

@article{Borsanyi:2016nature20115,
  author  = {Bors{\'a}nyi, Szabolcs and Fodor, Zolt{\'a}n and Guenther, Jana and Kampert, Karl-Heinz and Katz, S{\'a}ndor D. and Kawanai, Taichi and Kov{\'a}cs, Tam{\'a}s G. and others},
  title   = {Calculation of the axion mass based on high-temperature lattice quantum chromodynamics},
  journal = {Nature},
  volume  = {539},
  number  = {7627},
  pages   = {69--71},
  year    = {2016},
  doi     = {10.1038/nature20115},
  eprint  = {1606.07494},
  archivePrefix = {arXiv},
  primaryClass  = {hep-lat}
}

@article{Athenodorou:2022JHEP10_197,
  author  = {Athenodorou, Andreas and Bonanno, Claudio and Bonati, Claudio and Clemente, Giuseppe and D'Angelo, Francesco and D'Elia, Massimo and Maio, Lorenzo and Martinelli, Guido and Sanfilippo, Francesco and Todaro, Antonino},
  title   = {Topological susceptibility of $N_f=2+1$ QCD from staggered fermions spectral projectors at high temperatures},
  journal = {Journal of High Energy Physics},
  volume  = {2022},
  number  = {10},
  pages   = {197},
  year    = {2022},
  doi     = {10.1007/JHEP10(2022)197},
  eprint  = {2208.08921},
  archivePrefix = {arXiv},
  primaryClass  = {hep-lat}
}

@article{Chen:2022PhysRevD106_074501,
  author  = {Chen, Yu-Chih and Chiu, Ting-Wai and Hsieh, Tung-Han},
  title   = {Topological susceptibility in finite temperature QCD with physical $(u/d,s,c)$ domain-wall quarks},
  journal = {Physical Review D},
  volume  = {106},
  pages   = {074501},
  year    = {2022},
  doi     = {10.1103/PhysRevD.106.074501},
  eprint  = {2204.01556},
  archivePrefix = {arXiv},
  primaryClass  = {hep-lat}
}

@article{Kotov:2025JHEP09_045,
  author  = {Kotov, A. Yu. and Lombardo, M. P. and Trunin, A.},
  title   = {Topological observables and $\theta$ dependence in high temperature QCD from lattice simulations},
  journal = {Journal of High Energy Physics},
  volume  = {2025},
  number  = {09},
  pages   = {045},
  year    = {2025},
  doi     = {10.1007/JHEP09(2025)045}
}

@article{Hertzberg:2008wr,
  author  = {Hertzberg, M. P. and Tegmark, M. and Wilczek, F.},
  title   = {Axion cosmology and the energy scale of inflation},
  journal = {Phys. Rev. D},
  volume  = {78},
  pages   = {083507},
  year    = {2008},
  doi     = {10.1103/PhysRevD.78.083507}
}

@article{Hamann:2009yf,
  author  = {Hamann, J. and Hannestad, S. and Raffelt, G. G. and others},
  title   = {Cosmological constraints on neutrino and axion hot dark matter},
  journal = {JCAP},
  volume  = {06},
  pages   = {022},
  year    = {2009},
  doi     = {10.1088/1475-7516/2009/06/022}
}

@article{Beltran:2006sq,
  author  = {Beltran, M. and Garcia-Bellido, J. and Lesgourgues, J.},
  title   = {Bounds on hot dark matter particles with entropic degrees of freedom},
  journal = {Phys. Rev. D},
  volume  = {75},
  pages   = {103507},
  year    = {2007},
  doi     = {10.1103/PhysRevD.75.103507}
}

@book{Dodelson:2003ft,
  author    = {Dodelson, S.},
  title     = {{Modern Cosmology}},
  publisher = {Academic Press},
  address   = {Amsterdam},
  year      = {2003},
  pages     = {196}
}

@article{Husdal:2016haj,
  author  = {Husdal, L.},
  title   = {On Effective Degrees of Freedom in the Early Universe},
  journal = {Galaxies},
  volume  = {4},
  pages   = {78},
  year    = {2016},
  doi     = {10.3390/galaxies4040078}
}

@article{1968Theory,
  author  = {Mclachlan, N. W.},
  title   = {Theory and Application of Mathieu Functions},
  journal = {The Mathematical Gazette},
  volume  = {52},
  year    = {1968}
}

@article{Zheng:2025wga,
    author = "Zheng, Ruifeng and Wei, Puxian and Yang, Qiaoli",
    title = "{Generation of axions and axion-like particles through mass parametric resonance induced by scalar perturbations in the early universe}",
    eprint = "2507.13127",
    archivePrefix = "arXiv",
    primaryClass  = "hep-ph",
    doi = "10.1088/1674-1137/adfa82",
    journal = "Chin. Phys.",
    volume = "49",
    number = "12",
    pages = "125108",
    year = "2025"
}

@misc{AxionLimits,
  author       = {Ciaran O'Hare},
  title        = {cajohare/AxionLimits: AxionLimits},
  month        = jul,
  year         = 2020,
  publisher    = {Zenodo},
  version      = {v1.0},
  doi          = {10.5281/zenodo.3932430},
  howpublished = {\url{https://cajohare.github.io/AxionLimits/}}
}

@article{Kofman:1997yn,
  author  = {Kofman, L. and Linde, A. D. and Starobinsky, A. A.},
  title   = {Towards the theory of reheating after inflation},
  journal = {Phys. Rev. D},
  volume  = {56},
  pages   = {3258},
  year    = {1997},
  doi     = {10.1103/PhysRevD.56.3258}
}

@article{Sikivie:2021trt,
  author  = {Sikivie, P. and Xue, W.},
  title   = {Resonant excitation of the axion field during the QCD phase transition},
  journal = {Phys. Rev. D},
  volume  = {105},
  pages   = {043533},
  year    = {2022},
  doi     = {10.1103/PhysRevD.105.043533}
}

@misc{Ahmad:2025,
      title={Resonant enhancement of axion dark matter decay}, 
      author={Bilal Ahmad and Yu-Ang Liu and Nick Houston},
      year={2025},
      eprint={2507.18508},
      archivePrefix={arXiv},
      primaryClass={hep-ph},
      url={https://arxiv.org/abs/2507.18508}, 
}

@article{Planck:2018vyg,
    author = "Aghanim, N. and others",
    collaboration = "Planck",
    title = "{Planck 2018 results. VI. Cosmological parameters}",
    eprint = "1807.06209",
    archivePrefix = "arXiv",
    primaryClass = "astro-ph.CO",
    doi = "10.1051/0004-6361/201833910",
    journal = "Astron. Astrophys.",
    volume = "641",
    pages = "A6",
    year = "2020",
    note = "[Erratum: Astron.Astrophys. 652, C4 (2021)]"
}

@article{Planck:2018jri,
  author        = {Akrami, Y. and others},
  title         = {Planck 2018 results: X. Constraints on inflation},
  collaboration = {Planck},
  journal       = {Astron. Astrophys.},
  volume        = {641},
  pages         = {A10},
  year          = {2020},
  doi           = {10.1051/0004-6361/201833887}
}

@article{Persic:1995ru,
    author = "Persic, Massimo and Salucci, Paolo and Stel, Fulvio",
    title = "{The Universal rotation curve of spiral galaxies: 1. The Dark matter connection}",
    eprint = "astro-ph/9506004",
    archivePrefix = "arXiv",
    reportNumber = "SISSA-60-95-A",
    doi = "10.1093/mnras/278.1.27",
    journal = "Mon. Not. Roy. Astron. Soc.",
    volume = "281",
    pages = "27",
    year = "1996"
}

@article{MukhanovEtAl1992,
  author       = {Mukhanov, V. F. and Feldman, H. A. and Brandenberger, R. H.},
  title        = {Theory of Cosmological Perturbations},
  journal      = {Physics Reports},
  volume       = {215},
  number       = {5--6},
  pages        = {203--333},
  year         = {1992},
  doi          = {10.1016/0370-1573(92)90044-Z},
  url          = {https://doi.org/10.1016/0370-1573(92)90044-Z}
}

@article{MaBertschinger1995,
  author       = {Ma, Chung-Pei and Bertschinger, Edmund},
  title        = {Cosmological Perturbation Theory in the Synchronous and Conformal Newtonian Gauges},
  journal      = {The Astrophysical Journal},
  volume       = {455},
  pages        = {7--25},
  year         = {1995},
  doi          = {10.1086/176550},
  url          = {https://doi.org/10.1086/176550},
  eprint       = {astro-ph/9506072},
  archivePrefix= {arXiv}
}

@article{kaya,
    author = "Kaya, Ali",
    title = "{Fluctuations of Quantum Fields in a Classical Background and Reheating}",
    eprint = "0909.2712",
    archivePrefix = "arXiv",
    primaryClass = "hep-th",
    doi = "10.1103/PhysRevD.81.023521",
    journal = "Phys. Rev. D",
    volume = "81",
    pages = "023521",
    year = "2010"
}

@article{Fauth,
    author = {Fauth, Gregor and Berges, J{\"u}rgen and Di Piazza, Antonino},
    title = "{Collisional strong-field QED kinetic equations from first principles}",
    eprint = "2103.13437",
    archivePrefix = "arXiv",
    primaryClass = "hep-ph",
    doi = "10.1103/PhysRevD.104.036007",
    journal = "Phys. Rev. D",
    volume = "104",
    number = "3",
    pages = "036007",
    year = "2021"
}
\clearpage
\renewcommand{\theequation}{A\arabic{equation}}
\setcounter{equation}{0} 
\onecolumngrid
\appendix
\begin{center}
  \textbf{\large Supplementary Material for Parametric-Resonance Production of QCD Axions}\\[.2cm]
  \vspace{0.05in}
\end{center}

\section{The Fourier-space equation of motion}
\label{sect:fourier}

To connect Eq.~(\ref{eq2341_std}) to the mode-by-mode evolution used earlier, we Fourier transform the field and perturbations using the convention 
\begin{equation}
\phi(\vec{x},t)=\int\!\frac{{\rm d}^3k}{(2\pi)^3}\,e^{i\vec{k}\cdot\vec{x}}\,\phi_{\vec{k}}(t).
\end{equation}
The same applies to $\Psi(\vec{x},t)$ and $\delta T(\vec{x},t)$. Substituting into Eq.~(\ref{eq2341_std}), the homogeneous part becomes diagonal in $\vec{k}$:  
\begin{equation}
\ddot\phi_{\vec{k}}+3H\dot\phi_{\vec{k}}+\left(\frac{k^{2}}{a^{2}}+m^{2}(T(t))\right)\phi_{\vec{k}}
+ f_{\vec{k}}(t)=0~,
\label{eq_mode_master}
\end{equation}
where $k \equiv |\vec{k}|$ and $f_{\vec{k}}$ is the Fourier transform of $f(\vec{x},t,\phi)$. Since $f$ contains products of perturbations with $\phi$ and its derivatives, its Fourier transform generally involves a convolution over intermediate momenta~\cite{Bracewell1966TheFT}:  

\begin{equation}
f_{\vec{k}}(t)=\int\!\frac{{\rm d}^3q}{(2\pi)^3}\,\mathcal{K}(\vec{k},\vec{q};t)\,,
\label{eq_convolution_generic}
\end{equation}
The kernel $\mathcal{K}$ is obtained by transforming each term in $f$. Explicitly,
\begin{equation}
\begin{aligned}
f_{\vec{k}}(t)=
&-2\!\int\!\frac{{\rm d}^3q}{(2\pi)^3}\,\Psi_{\vec{k}-\vec{q}}(t)\,\ddot\phi_{\vec{q}}(t)
-\!\int\!\frac{{\rm d}^3q}{(2\pi)^3}\Big(\dot\Psi_{\vec{k}-\vec{q}}-3\dot\Phi_{\vec{k}-\vec{q}}+6H\Psi_{\vec{k}-\vec{q}}\Big)\dot\phi_{\vec{q}}+\frac{2}{a^2}\!\int\!\frac{{\rm d}^3q}{(2\pi)^3}\,\Phi_{\vec{k}-\vec{q}}(t)\,(-q^2)\,\phi_{\vec{q}}(t)
-\frac{1}{a^2}\\&\!\int\!\frac{{\rm d}^3q}{(2\pi)^3}\,i(\vec{k}-\vec{q})\!\cdot\! i\vec{q}\,\big(\Phi_{\vec{k}-\vec{q}}+\Psi_{\vec{k}-\vec{q}}\big)\phi_{\vec{q}}(t)+\frac{{\rm d}m^2}{{\rm d}T}\!\int\!\frac{{\rm d}^3q}{(2\pi)^3}\,\delta T_{\vec{k}-\vec{q}}(t)\,\phi_{\vec{q}}(t)~.
\end{aligned}
\label{eq_fk_full}
\end{equation}

Equations~(\ref{eq_mode_master}) and~(\ref{eq_fk_full}) are the exact Fourier-space form of Eq.~(\ref{eq2341_std}). In general, the products of inhomogeneous perturbations with the axion field generate convolution integrals and therefore couple different axion momenta. As in standard gradient-expansion and mean-field/Hartree closures,\cite{Fauth,kaya}.  In this treatment, $\Phi$, $\Psi$, and $\delta T$ are taken to be prescribed linear acoustic backgrounds, and their action on a given axion mode is approximated by effective time-dependent coefficients multiplying that same mode.  We neglect explicit inter-mode momentum transfer,  while preserving the leading linear resonant forcing that governs the per-mode amplification studied in the Letter~\cite{MukhanovEtAl1992,MaBertschinger1995}.

Accordingly, Eq.~\ref{foeq} should be interpreted as an effective open-subsystem equation for the axion sector: mode-by-mode axion energy need not be conserved during resonance, since the injected energy is supplied by the prescribed perturbed radiation/gravitational background, whereas exact energy-momentum conservation applies only to the full coupled system represented by Eqs.~(\ref{eq_mode_master}) and~(\ref{eq_fk_full})~.

In the regime studied here, Eq.~(\ref{foeq}) is solved numerically without further approximation, and the induced response remains linear, with $\phi_1\propto\Phi_P$ and $\rho_1\propto\Phi_P^2$ throughout the parameter range shown. The ensemble-averaged evolution within numerical uncertainty therefore indicates that explicit coupled-mode and lattice effects are subdominant for the relic-density correction in the linear regime analyzed here.
\section{Full equation of a Mathieu form}
\label{sect:mathieu_derivation}

Writing Eq.~\eqref{foeq} schematically:
\begin{equation}
\begin{split}
\Bigl[1 - \alpha(t)\cos\Theta\Bigr] \ddot{\phi}_k &+ \Bigl[\gamma_0(t) + \gamma_1(t) \sin\Theta + \gamma_2(t) \cos\Theta\Bigr] \dot{\phi}_k + \Bigl[\omega_k^2(t) + \delta\omega^2(t) \cos\Theta\Bigr] \phi_k = 0,
\end{split}
\label{eq:schematic_app}
\end{equation}
with coefficients:
\begin{eqnarray}
\alpha(t) &=& \frac{9\Phi_p}{2k^2 t t_1}, \quad \gamma_0(t) = \frac{3}{2t},\nonumber\\
\omega_k^2(t) &=& \frac{k^2}{a^2(t)} + m^2(t) = k^2 \frac{t_1}{t} + m^2(t), \\
{\rm and}~\delta\omega^2(t) &=& \frac{9\Phi_p}{2t^2} - \frac{3\Phi_p}{2} \frac{{\rm d}m^2(t)}{{\rm d}\ln T}.\nonumber
\end{eqnarray}
Near the resonant time $t \approx t_R(k)$, we assume a narrow window $\Delta t$ where:  
\begin{equation}
\Delta t \ll t_R, \quad \left|\frac{\dot{\omega_k}}{\omega_k}\right| \Delta t \ll 1, \quad \left|\frac{\dot{\delta\omega^2}}{\delta\omega^2}\right| \Delta t \ll 1,
\end{equation}
This allows treating $\omega_k^2(t)$ and $\delta\omega^2(t)$ as approximately constant while trigonometric terms oscillate rapidly. Neglecting Hubble friction and small modulations, Eq.~(\ref{eq:schematic_app}) reduces to:
\begin{equation}
\ddot{\phi}_k + \Bigl[\omega_k^2(t_R) + \delta\omega^2(t_R) \cos\Theta_k(t)\Bigr] \phi_k \simeq 0.
\end{equation}
Define a rescaled time variable:
\begin{eqnarray}
z \equiv \frac{\Theta_k(t)}{2}=\frac{k}{\sqrt{3}}\sqrt{t_1 t}=\frac{k}{2\sqrt{3}}\eta(t),
\qquad\Rightarrow\qquad
\cos\Theta_k(t)=\cos(2z).
\label{eq:zdef_app}
\end{eqnarray}
Within the resonant window, $\dot z$ is approximately constant. Using $\dot{\phi} = \dot{z}\,\phi'$ and $\ddot{\phi} \simeq \dot{z}^2\,\phi''$ (primes denote $d/dz$), we obtain:  
\begin{equation}
\phi_k'' + \left[\frac{\omega_k^2(t_R)}{\dot{z}^2} + \frac{\delta\omega^2(t_R)}{\dot{z}^2} \cos(2z)\right] \phi_k \simeq 0.
\end{equation}
Comparing to the canonical Mathieu equation:  
\begin{equation}
\phi_k'' + \Bigl(A_k - 2q \cos(2z)\Bigr) \phi_k = 0,
\end{equation}
we identify parameters:
\begin{equation}
A_k(t_R) = \frac{\omega_k^2(t_R)}{\dot{z}^2}, \quad -2q(t_R) = \frac{\delta\omega^2(t_R)}{\dot{z}^2}.
\end{equation}

Using $\omega_k^2 = k^2/a^2 + m^2$ and $\dot{z}^2 = k^2/(12a^2)$, we recover the standard Mathieu parameters $A_k$ and $q$. For small $|q|$, instability tongues emerge near $A_k \simeq \ell^2$, with the resonance centers at $A_k(t_R) = \ell^2$. Under a standard rescaling (where the Mathieu parameter $A_k$ is normalized by 4), the dominant instability band corresponds to $\ell = 2$, centered at $A_k \simeq 4$. This reflects a time-local commensurability between the oscillator frequency and the driving term:  
\begin{equation}
\begin{split}
A_k(t_R) \simeq 4 \quad \Longleftrightarrow \quad \frac{m(t_R) a(t_R)}{k} \simeq \frac{1}{\sqrt{3}} \Longleftrightarrow \quad k_{\rm phys}(t_R) \simeq \sqrt{3} m(t_R).
\label{2}
\end{split}
\end{equation}

For a power-law mass model $m(T) \propto T^{-n}$, this condition reduces to the resonance criterion:  
\medskip
\emph{Case 1: Resonance during growth ($t_R<t_2$).}
Substitute $m(t_R)=m(t_R/t_2)^n$ into Eq.~\eqref{2}:
\begin{equation}
k^2\frac{t_1}{t_R}=3m^2\left(\frac{t_R}{t_2}\right)^{2n}
\quad\Rightarrow\quad
t_R^{2n+1}=\frac{k^2 t_1 t_2^{2n}}{3m^2}.
\end{equation}
Define
\begin{equation}
k_2 \equiv m\sqrt{\frac{3t_2}{t_1}},
\label{eq:k2def_app}
\end{equation}
which gives
\begin{equation}
t_R=t_2\left(\frac{k}{k_2}\right)^{\frac{2}{2n+1}}\qquad (k<k_2).
\label{eq:tR_case1_app}
\end{equation}

\emph{Case 2: Resonance after saturation ($t_R>t_2$).}
Set $m(t_R)=m$ in Eq.~\eqref{2}:
\begin{equation}
k^2\frac{t_1}{t_R}=3m^2
\quad\Rightarrow\quad
t_R=\frac{k^2 t_1}{3m^2}=t_2\left(\frac{k}{k_2}\right)^2\qquad (k>k_2),
\label{eq:tR_case2_app}
\end{equation}

this matches the main text’s derived condition, validating consistency across both regimes.  

\end{document}